%% file: main.tex
\documentclass[letterpaper]{article} % DO NOT CHANGE THIS
\usepackage{aaai2026}  % DO NOT CHANGE THIS
\usepackage{times}  % DO NOT CHANGE THIS
\usepackage{helvet}  % DO NOT CHANGE THIS
\usepackage{courier}  % DO NOT CHANGE THIS
\usepackage[hyphens]{url}  % DO NOT CHANGE THIS
\usepackage{graphicx} % DO NOT CHANGE THIS
\usepackage{natbib}  % DO NOT CHANGE THIS AND DO NOT ADD ANY OPTIONS TO IT
\usepackage{caption} % DO NOT CHANGE THIS AND DO NOT ADD ANY OPTIONS TO IT
\usepackage{algorithm}
\usepackage{algorithmic}
\floatname{algorithm}{Protocol}

\usepackage{amssymb}  % For \mathbb
\usepackage{amsmath}

\usepackage{booktabs}
\usepackage{array}
\usepackage{multirow}
\usepackage{bbding} 
\usepackage{makecell}

\usepackage{stmaryrd} % 用于\hecipher
\usepackage{xspace}
\usepackage{amsthm}
\theoremstyle{definition}
\newtheorem{definition}{Definition}
\newcommand{\ashare}[1]{\langle {#1} \rangle}
\newcommand{\hecipher}[1]{\ensuremath{[\![#1]\!]}\xspace}

\newcommand{\SecMoE}{\ensuremath{\mathsf{SecMoE}}\xspace}
\usepackage{newfloat}
\usepackage{listings}

\usepackage{subcaption} 
\usepackage{enumitem}

\DeclareCaptionStyle{ruled}{labelfont=normalfont,labelsep=colon,strut=off} % DO NOT CHANGE THIS
\floatstyle{ruled}
\newfloat{listing}{tb}{lst}{}
\floatname{listing}{Listing}

\title{SecMoE: Communication-Efficient Secure MoE Inference via Select-Then-Compute}
\author{
    Bowen Shen\textsuperscript{\rm 1,\rm 2},
    Yuyue Chen\textsuperscript{\rm 1},
    Peng Yang\textsuperscript{\rm 1},
    Bin Zhang\textsuperscript{\rm 2$^*$},
    Xi Zhang\textsuperscript{\rm 3},
    Zoe L. Jiang\textsuperscript{\rm 1,\rm 4}\thanks{Corresponding authors.}
}
\affiliations{
    \textsuperscript{\rm 1}School of Computer Science and Technology, Harbin Institute of Technology, Shenzhen, Shenzhen, China\\
    \textsuperscript{\rm 2}Department of New Networks, Pengcheng Laboratory, Shenzhen, China\\
    \textsuperscript{\rm 3}College of Management and Economics, Tianjin University, Tianjin, China\\
    \textsuperscript{\rm 4}Guangdong Key Laboratory of New Security and Intelligence Technology, Shenzhen, China\\
    \textsuperscript{}{\{shenbowen, chenyuyue, stuyangpeng\}}@stu.hit.edu.cn, bin.zhang@pcl.ac.cn, jackyzhang@tju.edu.cn, zoeljiang@hit.edu.cn
}

\makeatletter
\@ifundefined{isChecklistMainFile}{
  \newif\ifreproStandalone
  \reproStandalonetrue
}{
  \newif\ifreproStandalone
  \reproStandalonefalse
}
\makeatother

\ifreproStandalone

\usepackage{times}
\usepackage{helvet}
\usepackage{courier}
\usepackage{xcolor}

\usepackage{makecell}
\newcommand{\valp}[2]{\makecell{$#1$\\(#2$\times$)}} % 上行主值，下行括号值

\begin{document}

\maketitle

\begin{abstract}
Privacy-preserving Transformer inference has gained attention due to the potential leakage of private information. Despite recent progress, existing frameworks still fall short of practical model scales, with gaps up to a hundredfold. A possible way to close this gap is the Mixture of Experts (MoE) architecture, which has emerged as a promising technique to scale up model capacity with minimal overhead. However, given that the current secure two-party (2-PC) protocols allow the server to homomorphically compute the FFN layer with its plaintext model weight, under the MoE setting, this could reveal which expert is activated to the server, exposing token-level privacy about the client's input. While naively evaluating all the experts before selection could protect privacy, it nullifies MoE sparsity and incurs the heavy computational overhead that sparse MoE seeks to avoid. To address the privacy and efficiency limitations above, we propose a 2-PC privacy-preserving inference framework, \SecMoE. Unifying per-entry circuits in both the MoE layer and piecewise polynomial functions, \SecMoE obliviously selects the extracted parameters from circuits and only computes one encrypted entry, which we refer to as Select-Then-Compute. This makes the model for private inference scale to 63$\times$ larger while only having a 15.2$\times$ increase in end-to-end runtime. Extensive experiments show that, under 5 expert settings, \SecMoE lowers the end-to-end private inference communication by 1.8$\sim$7.1$\times$ and achieves 1.3$\sim$3.8$\times$ speedup compared to the state-of-the-art (SOTA) protocols.
\end{abstract}

\section{Introduction}
% 引入安全多方计算（MPC）作为一种保护隐私的计算范式，强调其在分布式环境中保护数据和模型隐私的能力。
% 简述MoE（Mixture of Experts）架构在大模型中的作用，说明其通过专家网络和门控机制提升模型容量和计算效率的优势。
% 提出本论文的主要贡献：
% 首个使用MPC实现MoE架构大模型的隐私推理方案。
% 优化线性层的计算，通过同态密文乘法优化提升效率。
% 优化非线性层的计算，采用分段多项式近似逼近非线性函数。
% 优化MoE层中的TopK排序算法，提升安全计算性能。
% 概述论文结构，简要介绍后续章节的内容安排。

% 介绍大模型（如Transformer）在自然语言处理、图像识别等领域的广泛应用及其重要性。
Transformers~\cite{vaswani2017attention} have significantly enhanced machine learning capabilities across a range of tasks. In model scaling, the Mixture of Experts (MoE) architecture has emerged as a powerful technique in Transformer-based models, particularly in natural language processing~\cite{DBLP:journals/neco/JacobsJNH91}. By dynamically selecting a subset of experts for each input token, MoE significantly reduces computational overhead while maintaining high model capacity. 

% 指出大模型推理过程中面临的隐私挑战
However, despite great advantages, deploying the MoE models in an untrusted environment raises privacy concerns. On the one hand, the server that owns the model weights naturally expects privacy protection of its model weight because training a MoE model requires significant investments in both financial and computational resources. On the other hand, MoE inference requires clients to upload their prompts, which may contain sensitive user data, such as personal health records and biometric information. As a result, the server requires that client learn nothing about the model parameters beyond the inference outputs, and the client requires that the server learn nothing about client's input.

% 现有保护方案
To ensure privacy for Transformer inference, several works have focused on the realm of Privacy-Preserving Machine Learning (PPML), using secure multiparty computation (MPC)~\cite{wagh2018securenn,srinivasan2019delphi,li2022mpcformer,huang2022cheetah,rathee2020cryptflow2,mohassel2018aby3,hao2022iron,merge,mpcvit,dong2023puma, pang2023bolt,lu2023bumblebee,li2024nimbus, kei2025shaft}. However, those studies are mainly designed for relatively small-scale Transformer models, like BERT~\cite{kenton2019bert} and GPT-2~\cite{cohen2020gpt2}. Some works~\cite{dong2023puma,lu2023bumblebee} also test on larger models like LLaMA-7B~\cite{touvron2023llama}, but there is still a certain gap with the scale of parameters implemented in plaintext. Moreover, simply increasing the number of Transformer blocks to scale up the parameters leads to linearly growing performance overheads in MPC-based implementations.

% 现有方案的问题
To solve the scalability problems mentioned above, it is essential to develop privacy-preserving protocols for model scaling. Previous works~\cite{hao2022iron,pang2023bolt,lu2023bumblebee,li2024nimbus} compute linear layers with homomorphic encryption with the plaintext weight matrix held by the server. While sparse MoE requires the activation of only $k$ experts per token, and when plaintext expert weights are chosen before computation, the server will learn which weight matrices are accessed, leaking the client's private token information. The simple idea is to obliviously select the experts after evaluating all the experts securely with plaintext weight matrices. However, this method will nullify the MoE layer's sparsity, incurring the heavy computational overhead that the sparse MoE seeks to avoid.

% 主图
\begin{figure*}
    \centering
    \includegraphics[width=0.95\linewidth]{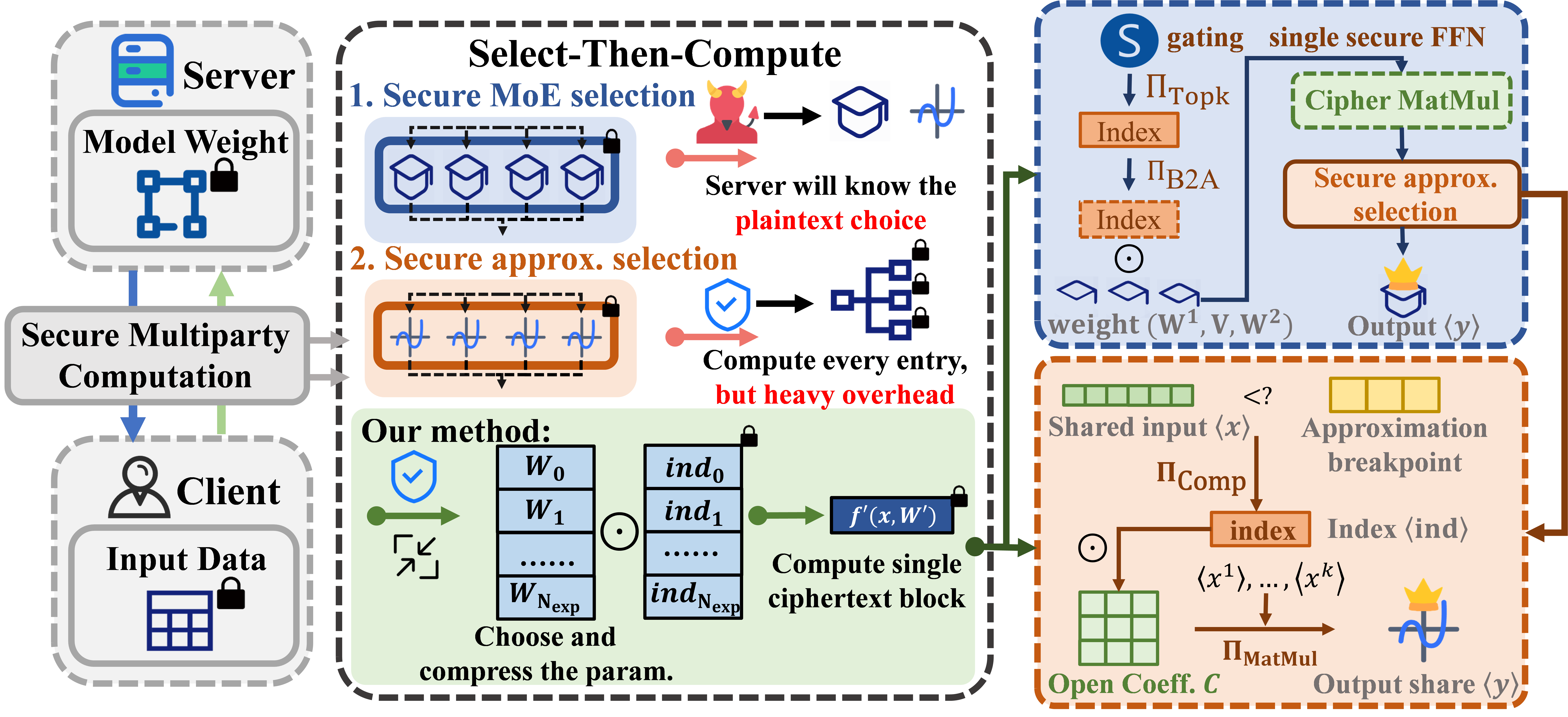}
    \caption{An overview of \SecMoE, deploying Select-Then-Compute in Secure MoE FFN and secure approximation selection.}
    \label{fig:framework}
\end{figure*}
% 提出自己的想法
To address these privacy and efficiency problems, in this work, we adopt the MoE architecture to serve as the foundation of our approach. As mentioned above, naively deploying existing 2-PC protocols to sparse MoE creates new privacy issues. Therefore, we propose a novel privacy-preserving inference framework for MoE architectures via Select-Then-Compute, named \SecMoE. Select-Then-Compute, including the selection phase and the compute phase, works on the functions that have multiple computing entries. The selection phase modifies each computing entry until all entries share the same computing circuit and extracts each entry's parameters as the choices. Next, the parameters will be obliviously selected by the ciphertext vector. Then, in the compute phase, only one encrypted entry will be computed with the chosen parameter. We design the secure sparse MoE and secure polynomial selection using Select-Then-Compute, as illustrated in Figure~\ref{fig:framework}. Specifically, in the MoE layer, our approach leverages lattice-based additive homomorphic encryption, combined with carefully designed selection bit strings, to implement secure sparse MoE. We encrypt the expert selection process in such a way that only the chosen experts' computations are performed, while keeping the selection private from the server. To further boost the model efficiency, we also form the nonlinear function piecewise polynomial approximation into the Select-Then-Compute. Specifically, we transform piecewise polynomials by padding all segments into uniform computing entries. Then, entries are obliviously aggregated by the encrypted selection vector so that only the activated segment is computed in the ciphertext. 

In summary, we propose a new secure 2-PC inference framework \SecMoE, and contributions are summarized as:
\begin{itemize}[nosep]
    \item We propose a new secure sparse MoE layer via Select-Then-Compute. Our protocol obliviously selects the extracted experts' parameter in the selection phase and obtains output through single expert evaluation in the compute phase. This preserves the MoE's sparsity and achieves almost linear communication with the increasing number of experts.
    
    \item For nonlinear layers, we redesign secure piecewise polynomial evaluation via Select-Then-Compute. Our protocol secretly selects the open coefficient matrix and performs low-degree polynomial evaluation in ciphertext. This achieves lower error in accuracy and higher efficiency in communication.
    
    \item % 补充先选后算的思想
    We combine the protocols above as \SecMoE and evaluate the performance using 2 MoE models in 5 expert settings under both LAN and WAN networks. Compared with the SOTA PPML frameworks Iron\cite{hao2022iron} and BumbleBee\cite{lu2023bumblebee}, \SecMoE reduces the communication of the private inference by 1.8$\sim$29.8$\times$ and improves end-to-end performance by 1.3$\sim$16.1$\times$.
 \end{itemize}

\section{Preliminaries}
\subsection{Threat Model}
% 半诚实设置
Our framework operates in a two-party setting, where the client $C$ possesses a private input, and the server $S$ holds the model weight. It ensures security against a semi-honest adversary, where both parties follow the protocol’s rules but attempt to extract unauthorized information from the protocol. Many PPML works use same setting\cite{juvekar2018gazelle,lu2023bumblebee}, and details refer to the Appendix.

\subsection{Notation}
Let $\mathbb{Z}_{2^{\ell}}$ denote the ring of integers modulo $2^{\ell}$. 
For any positive integer $n$, let $[n]=\{0,1,\dots,n-1\}$ denote its index set. 
For the homomorphic encryption (HE), 
we define $\mathbb{A}_{N,2^{\ell}} = \mathbb{Z}_{2^{\ell}}[X]\big/\!\bigl(X^{N}+1\bigr),$ 
where $N$ is an integer with a power of two. The elements in $\mathbb{A}_{N,2^{\ell}}$ 
are polynomials of degree at most $N-1$. And $\hecipher{\text{M}}$ 
denotes the homomorphic encryption ciphertext with its plaintext $\text{M}$. 
In the MoE Layer, we use $\text{W}_{i}$ to represent the plain weight matrix in the $i$-th expert. 
Additive arithmetic secret shares used in our work are represented as $\langle\cdot\rangle$ and 
boolean shares as $\langle\cdot\rangle^b$. $\Pi_{\text{MUX}}(\ashare{a}^b, \ashare{b})$ 
represents $\ashare{a?b:0}$. To illustrate clearly, 
we set $\Pi_{\text{Mul}}(a,b)=\Pi_{\text{trunc}}(\Pi_{\text{MUL}}(a,b))$, where $\Pi_{\text{MUL}}$ and $\Pi_{\text{trunc}}$ represent 2-PC arithmetic multiplication and secure truncation protocol followed by BumbleBee~\cite{lu2023bumblebee}.

\subsection{Cryptographic Primitives}

\paragraph{Additive Secret Sharing.}
% 加性秘密分享
In our two-party setting, the nonlinear layers are implemented through the additive secret sharing over the ring $\mathbb{Z}_{2^{\ell}}$. For example, to secretly share the client $C$'s value $x\in\mathbb{Z}_{2^{\ell}}$, $C$ uniformly samples a random share $\ashare{x}_s\in\mathbb{Z}_{2^{\ell}}$ and sends it to server $S$ as its secret share. $S$ sets its own share as $\ashare{x}_c = x-\ashare{x}_s$. The point-wise secure multiplication with shares needs oblivious transfer\cite{yang2020ferret}-based Beaver Triples\cite{beaver1996correlated}.

\paragraph{Lattice-based Additive Homomorphic Encryption.}
We use the lattice-based additive HE scheme~\cite{lu2023bumblebee} to build our linear layers. 
Homomorphic encryption allows one party to perform computations on the encrypted data of the other party without the need for the decryption key. This HE scheme~\cite{lu2023bumblebee} encodes a plaintext vector ${x} \in (\mathbb{Z}_{2^\ell})^N$ into a plaintext polynomial $\hat{x} \in \mathbb{A}_{N, 2^{\ell}}$, and then $\hat{x}$ is encrypted to a ciphertext $\text{Enc}(\hat{x})={\hecipher{x}} \in \mathbb{A}_{N, q}^2$ where $q$ is a ciphertext modulus. Therefore, the inputs in our linear layers are $\ashare{x}_c, \ashare{x}_s \in \mathbb{Z}_{2^{\ell}}$ with dimension $k\times m$. And sever holds the plaintext weight $\text{W}\in \mathbb{Z}_{2^{\ell}}$ with dimension $m\times n$. The encoding functions $\pi_{L}$ and $\pi_{R}$ will encode inputs into polynomials $\hat{x}$ and $\hat{w}$: $\hat{x}=\pi_{L}(x):\hat{x}\!\bigl[i m n + (m-1) - j\bigr]= x_{i,j}, i \in [k], j \in [m], \hat{w}= \pi_{R}(\text{W}):\hat{w}\!\bigl[j m + i\bigr] = \text{W}_{i,j}, i \in [m], j \in [n].$

% \begin{equation}
% \begin{aligned}
% \hat{x}
%   &= \pi_{L}(x):
%      \hat{x}\!\bigl[i m n + (m-1) - j\bigr]
%        = x_{i,j},
%       i \in [k], j \in [m], \\
% \hat{w}
%   &= \pi_{R}(\text{W}):
%      \hat{w}\!\bigl[j m + i\bigr]
%        = \text{W}_{i,j},
%      i \in [m], j \in [n].
% \end{aligned}
% \label{eq:bumblebee}
% \end{equation}

\subsection{Transformer Architecture}
\paragraph{Attention Layer.}
The attention Layer~\cite{vaswani2017attention} can be described as mapping a query key $\text{X}_\text{Q}$ and a set of key-value pairs ($\text{X}_\text{K}$,$\text{X}_\text{V}$) to a weighted sum, where the weights are calculated from the query key and the corresponding key values, and the formal expression is as shown in the following formula: $\text{Attention}(\text{X}_\text{Q},\text{X}_\text{K},\text{X}_\text{V}) = \text{Softmax}(\frac{\text{X}_\text{Q} \text{X}_\text{K}^\text{T}}{\sqrt \text{d}})\text{X}_\text{V},$
% \begin{equation}
%     \text{Attention}(\text{X}_\text{Q},\text{X}_\text{K},\text{X}_\text{V}) = \text{Softmax}(\frac{\text{X}_\text{Q} \text{X}_\text{K}^\text{T}}{\sqrt \text{d}})\text{X}_\text{V},
% \label{eq:1}
% \end{equation}
where $d$ is the hidden dimension. $\text{X}_\text{Q}$, $\text{X}_\text{K}$, and $\text{X}_\text{V}$ are different linear projections of the input, satisfying: $\text{X}_\text{Q}=\text{W}_\text{Q}\cdot \text{X}, \text{X}_\text{K}=\text{W}_\text{K}\cdot \text{X}, \text{X}_\text{V}=\text{W}_\text{V}\cdot \text{X},$ $\text{W}_\text{Q}, \text{W}_\text{K}, \text{W}_\text{V}$ are the pre-trained weight matrices. Multi-head attention extends the above mechanism to parallel attention layers.

\paragraph{FeedForward Network (FFN).}
Our work mainly focuses on FeedForward Network with GeGLU functions. It consists of two fully connected layers with an activation function in between:
\begin{equation}
\text{FeedForward}(x)=\text{W}^2 (\sigma(\text{W}^1 x) \otimes \text{V} x),
\end{equation}
where $\text{V}$, $\text{W}^1$ and $\text{W}^2$ are parameter matrices. As for the activation $\sigma$, we apply GeLU function. Most large language models still use ReLU in the FFN (e.g., Switch Transformer Base and Large~\cite{fedus2022switch}). GPT models replace it with the smoother activation GeLU and GeGLU. The LLaMA series goes further, adopting the gated SwiGLU activation to boost expressiveness at minimal extra cost~\cite{touvron2023llama}. 

\paragraph{Mixture-of-Experts (MoE).}
% MoE段首
In large language models employing transformer architectures, the mixture-of-experts (MoE) layer is composed of a collection of $N_{\text{exp}}$ expert sub-networks ${\text{FFN}_0,..., \text{FFN}_{N_{\text{exp}}-1}}$, complemented by a gating network $G$. The gating network $G$ allocates the input to the most suitable expert sub-networks~\cite{jacobs1991adaptive,jordan1994hierarchical,collobert2001parallel}. The MoE layer is strategically integrated to replace the FFN within each transformer block, reducing the computational complexity of the FFN with model scaling. Based on the strategy of gating network $G$, MoE layers can be categorized into two types: dense MoE and sparse MoE, and we mainly focus on the sparse MoE. 

Sparse MoE is proposed by Shazeer~\cite{shazeer2017outrageously}, selectively activating a subset of expert sub-networks during each block. By calculating a weighted aggregation of outputs from the top-k experts, sparse MoE reduces the significant computational overhead compared to dense MoE:
    \begin{equation}
    \text{M}_{\text{Sparse}}(x) = \sum_{i=1}^{N_{\text{exp}}} \text{Softmax}(\Pi_{\text{Topk}}\left(g(x))\right)_i \text{FFN}_i(x),
    \end{equation}
    \begin{equation}
        \Pi_{\text{Topk}}(g(x)_i) = 
        \begin{cases} 
            g(x)_i, & \text{if } g(x)_i \in g(x)^{K_{\text{exp}}}, \\
            -\infty, & \text{else}.
         \end{cases},
    \end{equation}
where $g(x)$ represents the input of the Softmax operation and $g(x)^{K_{\text{exp}}}$ is the top-k elements of it. The hyperparameter $K_{\text{exp}}$ is the number of experts being selected, which is set as 1 in our work.

\section{Secure Sparse MoE Layer}
In this section, we first analyze the privacy problem in the secure MoE layer. Then, we propose the secure sparse MoE protocol via Select-Then-Compute. Further, we form the secure GeLU as a Select-Then-Compute function and propose its design in sparse MoE. We also analyze complexity in both secure sparse MoE and secure GeLU. We defer the correctness and the security proofs to the Appendix.

\subsection{Secure Sparse MoE Protocol}
The sparse MoE and the dense MoE model differ significantly in terms of their computational strategies. While dense MoE applies all available experts during each inference pass, which we do not give special optimizations, sparse MoE is designed to dynamically select only a subset of experts based on the input~\cite{fedus2022switch}, thus ensuring computational efficiency and scalability. The primary advantage of sparse MoE is its ability to increase the number of model parameters significantly, without the corresponding increase in computational overhead, by activating only $K_{\text{exp}}$ experts using the gating network $G$.

\begin{figure}
    \centering
    \includegraphics[width=\linewidth]{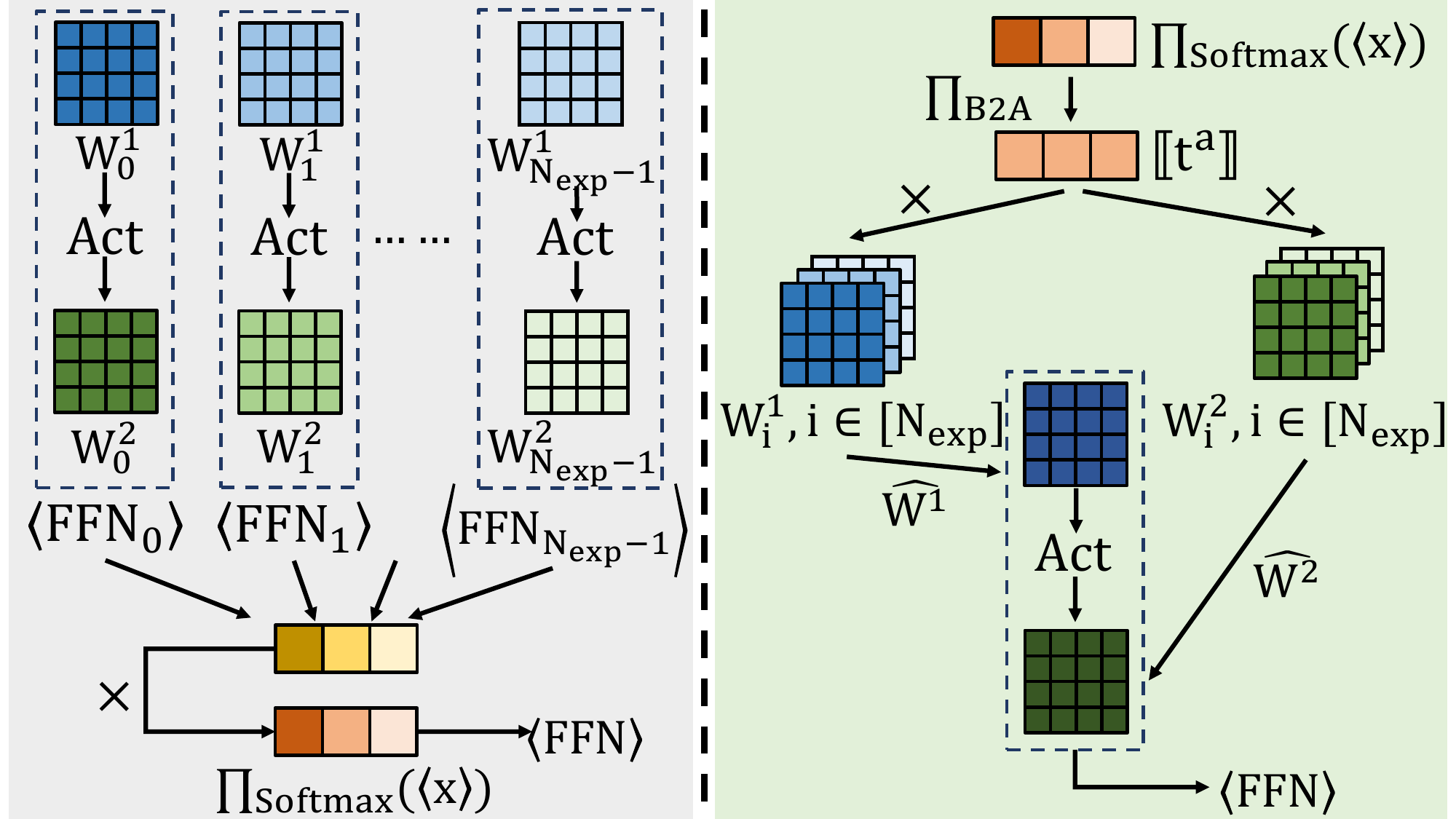}
    \caption{Illustration of secure MoE layer. Comparison between a naive privacy-preserving MoE (left) and our proposed secure sparse MoE (right). The baseline evaluates all experts and aggregates their outputs via Softmax, leading to high overhead. In contrast, our method securely selects top experts and only evaluates one expert, significantly reducing computation and communication cost.}
    \label{fig:moe}
\end{figure}

However, in the MPC setting, secure sparse MoE cannot be directly derived from secure dense MoE. If the secure dense MoE approach were adopted unchanged, all experts would be evaluated in parallel to maintain the expert selection oblivious. This strategy suffers significant inefficiency, as it leads to the unnecessary computation of all experts, which would incur at least additional $N_{\text{exp}}-K_{\text{exp}}$ experts' computational and communication overheads, which goes against the original intent of sparse MoE. To address the challenges above, we propose a modified approach to implement the secure sparse MoE via Select-Then-Compute, shown in Figure~\ref{fig:moe}, and we illustrate it in the next paragraph.

\paragraph{Select-Then-Compute in Secure Sparse MoE Protocol.}
In the secure sparse MoE Protocol, the client and server each hold additive secret shares of the input $\ashare{x}$, and the server additionally holds all experts' plaintext weight matrix sets $(\text{W}^1_i \in \mathbb{Z}_{2^{\ell}}^{m \times n}, \text{V}_i \in \mathbb{Z}_{2^{\ell}}^{m \times n}, \text{W}^2_i \in \mathbb{Z}_{2^{\ell}}^{n \times m}), i\in [N_{\text{exp}}]$. 

Select-Then-Compute has two phases: the selection phase and the compute phase. The entry in the secure sparse MoE protocol denotes each expert FFN. Noticed that each expert has the same components, and therefore, MoE experts are naturally unified. In the selection phase, \SecMoE leverages the advantage of homomorphic encryption's local communication-free advantages. It only requires sharing a $N_{\text{exp}}$-length selection vector, enabling oblivious selection of the encrypted weights. This significantly reduces online communication costs compared to previous works, which is shown in our experiments. Next, in the compute phase, \SecMoE computes similarly to previous work, with the key difference being that the output of the selection phase is a homomorphic ciphertext to prevent leaking choice to the server, whereas prior approaches~\cite{hao2022iron,li2024nimbus,lu2023bumblebee} typically used plaintext. While this introduces additional computational overhead of ciphertext-ciphertext matrix multiplication, as noted in Protocol~\ref{tab:secure_sparse_moe}, we only invoke this multiplication operator twice, making it more efficient compared to computing the remaining $N_{\text{exp}}-1$ expert FFNs. The whole protocol is described in Protocol~\ref{tab:secure_sparse_moe}.

\begin{algorithm}[!h]
   \caption{Secure Sparse MoE $\Pi_{\text{SparseMoE}}$}
   \label{tab:secure_sparse_moe}
\begin{algorithmic}[1]
        \renewcommand{\algorithmicrequire}{\textbf{Input:}}
        \REQUIRE The client $C$ and the server $S$ each holds $\ashare{x}_c$ and $\ashare{x}_s$, where $x=\ashare{x}_c+\ashare{x}_s$. The server holds the FFN weight matrices sets $(\text{W}^1_i \in \mathbb{Z}_{2^{\ell}}^{m \times n}, \text{V}_i \in \mathbb{Z}_{2^{\ell}}^{m \times n}, \text{W}^2_i \in \mathbb{Z}_{2^{\ell}}^{n \times m}), i\in [N_{\text{exp}}]$, where $m$, $n$ and $N_{\text{exp}}$ are the model, hidden dimension and number of experts, with the secret key $sk$.
        
        \renewcommand{\algorithmicrequire}{\textbf{Output:}}
        \REQUIRE $C$ and $S$ obtain $\ashare{y}_c$ and $\ashare{y}_s$ respectively, where $\Pi_{\text{SparseMoE}}(x)=\ashare{
        y}_c+\ashare{y}_s$.
    
    \COMMENT{\textbf{Selection Phase}} 
    \STATE $C$ and $S$ compute $\Pi_{\text{Topk}}$ to get top-k large values and output secret-shared index $\ashare{\text{SortVal}}$. 
    
    \STATE $C$ and $S$ compute $\Pi_{\text{onehot}}(\ashare{\text{SortVal}},K_{\text{exp}})$ outputs a secret-shared boolean vector $t^b \in \{0,1\}^{N_{\text{exp}}}$ with $t^b_{\text{index}}=1$, else 0. 
    
    \STATE $C$ and $S$ compute $\Pi_{\text{B2A}}$ converting the secret input $t^b$ from Boolean to arithmetic form (B2A) and outputs $t^a$. 

    \STATE $C$ encrypts $\hecipher{t^a_c}:=\text{Enc}(t^a_c)$ and sends to $S$. On receiving the cipher, $S$ computes $\hecipher{t^a}:=\hecipher{t^a_c}+t^a_s$.
    
    % \STATE For each expert's independent weight matrices $(\text{W}^1_i,\text{V}_i,\text{W}^2_i), i\in [K]$, $S$ homomorphically encrypt weight to obtain the corresponding ciphertexts $\hecipher{\text{W}^1}_i$, $\hecipher{\text{V}}_i$ and $\hecipher{\text{W}^2}_i$, $i\in[K]$. 
    
    \STATE $S$ computes point-wise secure multiplication $\Pi_{\text{Mul}}(\hecipher{t^a}, \text{W}^1_i)$ yielding $\hecipher{\text{W}^1_r} = \sum_{i=0}^{N_{\text{exp}}-1} \text{W}^1_i \cdot \hecipher{t^a}$, and similar for $\Pi_{\text{Mul}}(\hecipher{t^a},\text{V}_i),\Pi_{\text{Mul}}(\hecipher{t^a},\text{W}^2_i)$. 

    \COMMENT{\textbf{Compute Phase}}
    \STATE $C$ encrypts $\hecipher{\ashare{x}_c}:=\text{Enc}(\ashare{x}_c)$ and sends to $S$. After receiving, $S$ local computes $\hecipher{x}=\hecipher{\ashare{x}_c}+\ashare{x}_s$. 
    
    \STATE $S$ computes $\hecipher{\text{W}^1_r}\cdot\hecipher{x}$ and $\hecipher{\text{V}_r}\cdot\hecipher{x}$, and $\hecipher{\text{W}^1_r\cdot x-\text{R}^1}, \hecipher{\text{V}_r\cdot x-\text{R}^V}$ with random masks $\text{R}^1, \text{R}^V$. 
    
    \STATE $S$ sends $\hecipher{\text{W}^1_r\cdot x-\text{R}^1}$ and $\hecipher{\text{V}_r\cdot x-\text{R}^V}$ to $C$ and sets $(\ashare{x^{\text{GLU}}}_s, \ashare{x^V}_s)=(\text{R}^1, \text{R}^V)$. $C$ decrypts and obtains $(\ashare{x^{\text{GLU}}}_c, \ashare{x^V}_c)=(\text{W}^1_r\cdot x-\text{R}^1,\text{V}_r\cdot x-\text{R}^V)$. 
    
    \STATE $C$ and $S$ compute $\ashare{\text{act}}:=\Pi_{\text{GeLU}}(\ashare{x^{\text{GLU}}})$ and obtains  $\ashare{\text{act}}_c$ and $\ashare{\text{act}}_s$.
    
    \STATE $C$ and $S$ compute point-wise multiplication $\Pi_{\text{Mul}}(\ashare{\text{act}}_i,\ashare{x^V}_i)$ to obtain $\ashare{\text{GLU}}_c$ and $\ashare{\text{GLU}}_s$. 
    
    \STATE $C$ encrypts $\hecipher{\ashare{\text{GLU}}_c}:=\text{Enc}(\text{GLU}_c)$ and sends to $S$. $S$ locally computes $\hecipher{\text{GLU}}=\hecipher{\ashare{\text{GLU}}_c}+\ashare{\text{GLU}}_s$. 
    
    \STATE $S$ computes the cipher-cipher multiplication $\hecipher{\text{W}^2_r}\cdot\text{GLU}$ with $\hecipher{\text{GLU}}$ and $\hecipher{\text{W}^2_r}$. 
    
    \STATE $S$ computes $\hecipher{\text{W}^2_r\cdot\text{GLU}-\text{R}^2}$ with random $\text{R}^2$, and sends to $C$. $S$ sets $\text{R}^2$ as the $\ashare{y}_s$. 
    
    \STATE $C$ decrypts $\hecipher{\text{W}^2_r\cdot\text{GLU}-\text{R}^2}$ and sets $\text{W}^2_r\cdot\text{GLU}-\text{R}^2$ as its output share $\ashare{y}_c$. 
    
\end{algorithmic}
\end{algorithm}

\paragraph{Secure Sparse MoE Complexity Analysis.}
% Sparse
At the layer level, our approach eliminates redundant computation for at least $N_{\text{exp}}-1$ non‑selected experts by executing the selection phase, preserving sparsity under the MoE setting. The resulting overhead arises primarily from the ciphertext matrix multiplications in the compute phase. In practice, these costs are relatively small compared to the computation saved by redundant $N_{\text{exp}}-1$ inactive experts.

At the expert level, compared with BumbleBee, \SecMoE requires the conversion protocol $\Pi_{\text{B2A}}$ on the $N_{\text{exp}}$-dimensional one-hot vector with $cost_{\text{B2A}}$, the communication for $\hecipher{\ashare{t_c^a}}$ and local $\Pi_{\text{MatMul}}$. In the compute phase, unlike BumbleBee computes $\text{W}\cdot \hecipher{x}$, \SecMoE requires three cipher multiplications $\hecipher{\text{W}_r^1}\cdot \hecipher{x}$, $\hecipher{\text{V}_r}\cdot \hecipher{x}$, and $\hecipher{\text{W}_r^2}\cdot \hecipher{\text{GLU}}$. However, compared to the redundant computation of all experts dominating the costs, the additional cost introduced by ciphertext multiplication is fixed and relatively small. Therefore, this optimization becomes increasingly effective as the number of experts grows, by up to 29.8$\times$.

\subsection{Secure GeLU Protocol}
\paragraph{Select-Then-Compute in Secure Polynomial Evaluation.}
Previous works adopt piecewise polynomial approximation to calculate the GeLU function. Bumblebee~\cite{lu2023bumblebee} approximates the GeLU function with 4-piece polynomials at most degree-6. Nimbus~\cite{li2024nimbus} uses 3-piece polynomials at most degree-2, boosting further efficiency. However, these works primarily optimize polynomial-based expressions at the design stage, yet uniformly rely on the same secure polynomial evaluation paradigm during computation, illustrated in the Appendix. To address this inefficient communication problem, we describe the secure polynomial evaluation as a Select-Then-Compute process, and its entry denotes each segmented polynomial.

Before the Select-Then-Compute process, we first describe the preparation for polynomials. To form the approximation of the GeLU gate inside each expert, we follow the approximation interval $[-5,3]$ adopted by Bumblebee~\cite{lu2023bumblebee}. Within this interval, we fit piecewise quadratic polynomials using $\texttt{numpy.polynomial}$, solving a least‑squares problem in Chebyshev space to obtain the coefficients of two polynomials, achieving maximum absolute error of 1.2$\times 10^{-2}$ and mean absolute error of 1.7$\times 10^{-3}$ over the evaluation range. The values smaller than $-5$ are mapped to $0$, while values larger than $3$ are mapped to the $x$, as shown in Equation~\ref{eq:gelu}:

\begin{equation}
    \operatorname{GeLU}(x) = \begin{cases} 0 & x \in (-\infty,-5] \\ 
    P_1(x)    & x \in (-5,-3] \\ 
    P_2(x)    & x \in (-3,-1] \\
    P_3(x)    & x \in (-1,1] \\
    P_4(x)    & x \in (1,3] \\
    x           & x\in (3,\infty)
    \end{cases}
    \label{eq:gelu}
\end{equation}
and $P_1(x), P_2(x), P_3(x), P_4(x)$ are degree-2 polynomials. 

% \begin{equation}
%    \text{Coeff}\;=\;
%     \begin{bmatrix}
%     0.0 & 0.0 & 0.0 \\
%     c_{1,0} & c_{1,1} & c_{1,2}  \\
%     c_{2,0} & c_{2,1} & c_{2,2} \\
%     c_{3,0} & c_{3,1} & c_{3,2} \\
%     c_{4,0} & c_{4,1} & c_{4,2} \\
%     0.0 & 1.0 & 0.0
%     \end{bmatrix}
% \end{equation}

In the selection phase, to evaluate the piecewise polynomials in a pre-chosen manner, we first collect the polynomial coefficients into a matrix, where the row index $i\in\{1,\dots,m_{\text{seg}}\}$ enumerates the $m_{\text{seg}}$ segments and the column index $j\in\{1,\dots,n_{\text{seg}}\}$ enumerates the coefficients from the highest to the constant term. We pad each entry with 0 until it reaches the largest power among all segments. The resulting coefficient details are illustrated in the Appendix. Given an input $\ashare{x}$, we first compute secure comparisons between $\ashare{x}$ and each open plaintext breakpoint $b_i,i\in\{1,\dots,m_{\text{seg}}-1\}$, to obtain a boolean one‑hot segment selector. Therefore, in the selection phase, the hybrid coefficient row corresponding to $x$ can be retrieved by a single masked matrix–vector product. In the compute phase, given that we unify each entry power, the final value is then obtained by securely evaluating one polynomial with the largest power.
% 缺点：需要统一到分段中最高次幂，因此一次幂更优

\begin{figure*}[!t]
  \centering
  % 第1行
  \begin{subfigure}{0.32\textwidth}
    \includegraphics[width=\linewidth]{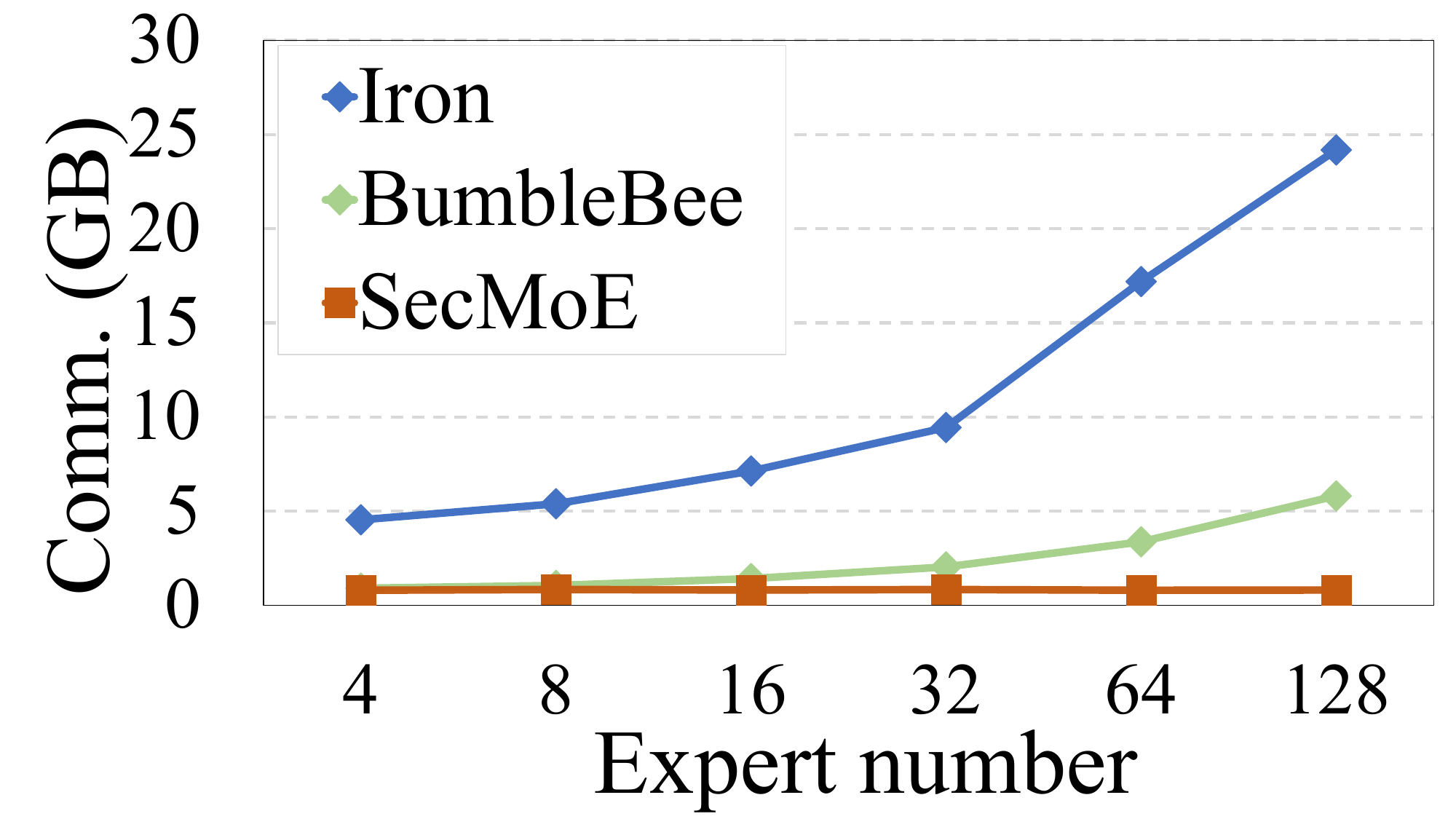}
  \end{subfigure}\hfill
  \begin{subfigure}{0.32\textwidth}
    \includegraphics[width=\linewidth]{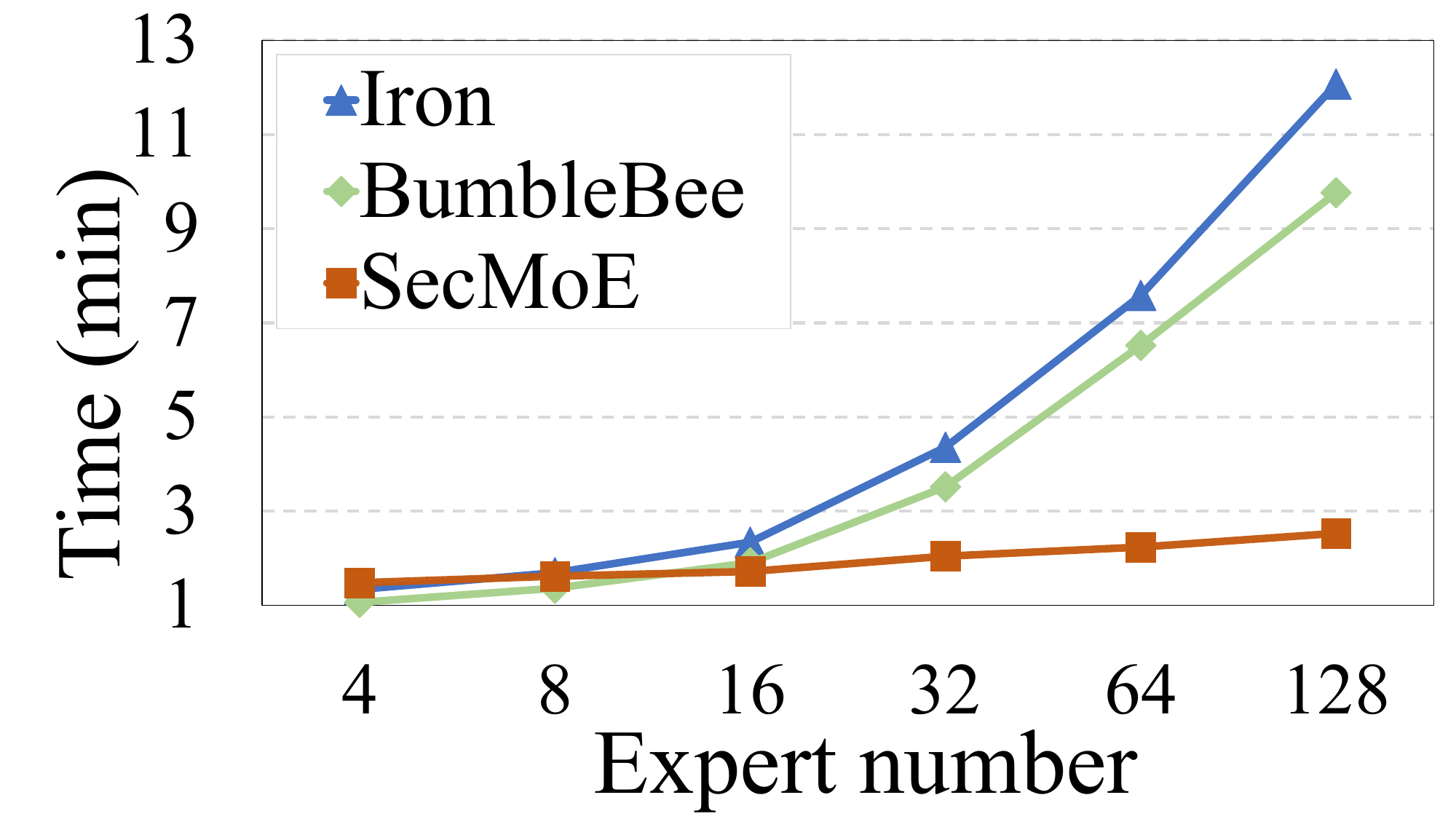}
  \end{subfigure}\hfill
  \begin{subfigure}{0.32\textwidth}
    \includegraphics[width=\linewidth]{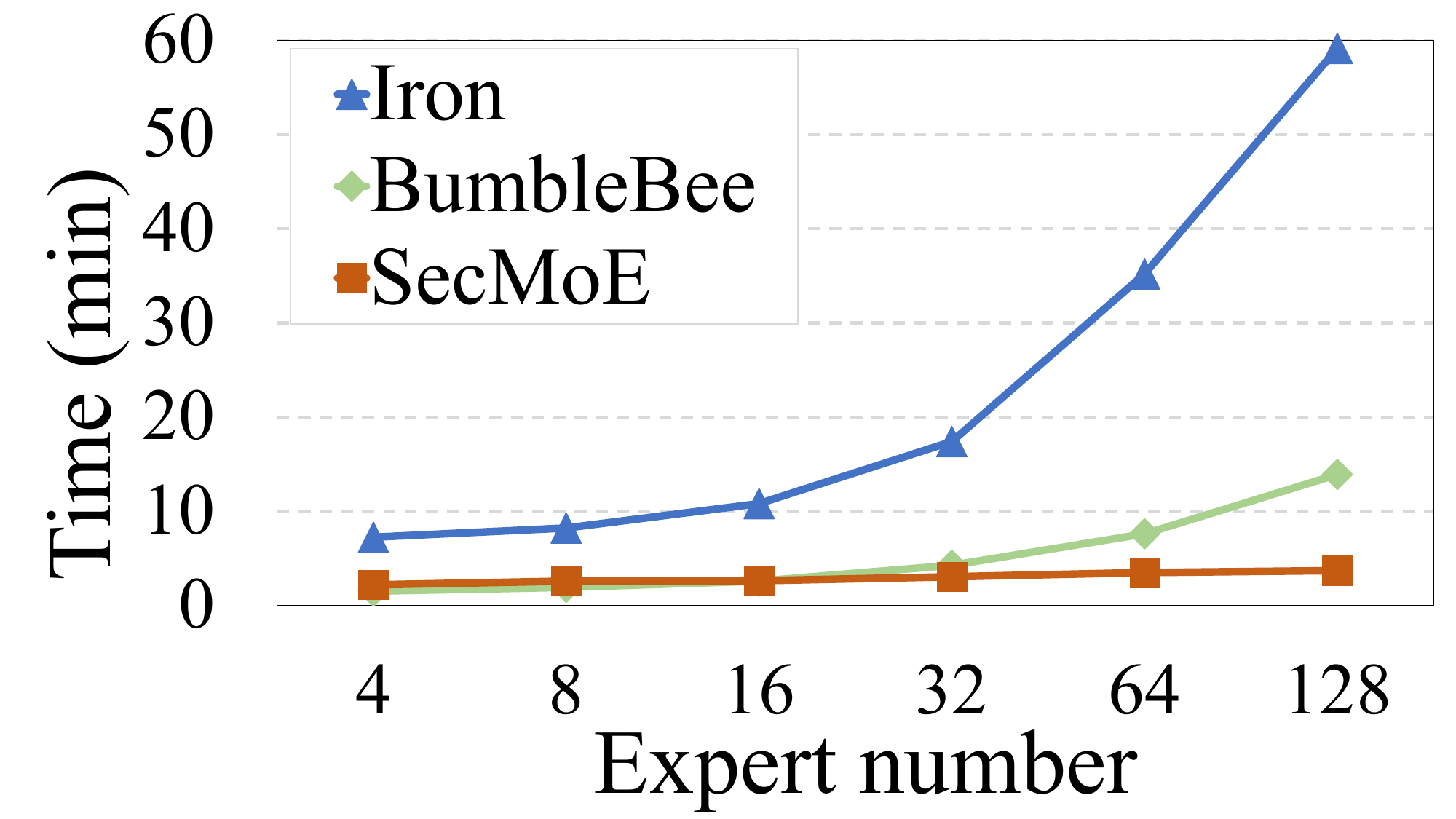}
  \end{subfigure}

  \medskip % 行间距（可用\vspace{4pt}微调）

  % 第2行
  \begin{subfigure}{0.32\textwidth}
    \includegraphics[width=\linewidth]{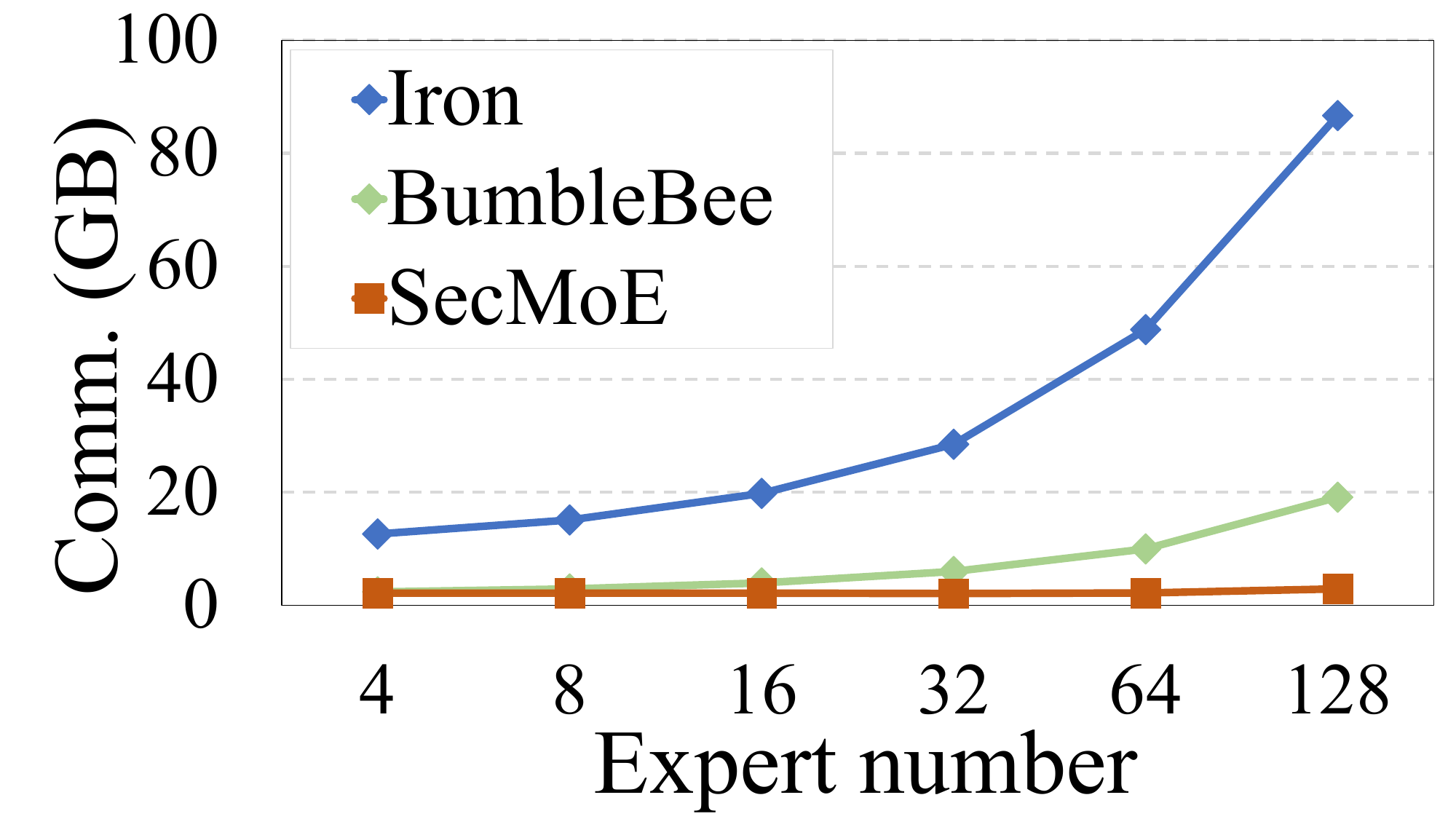}
  \end{subfigure}\hfill
  \begin{subfigure}{0.32\textwidth}
    \includegraphics[width=\linewidth]{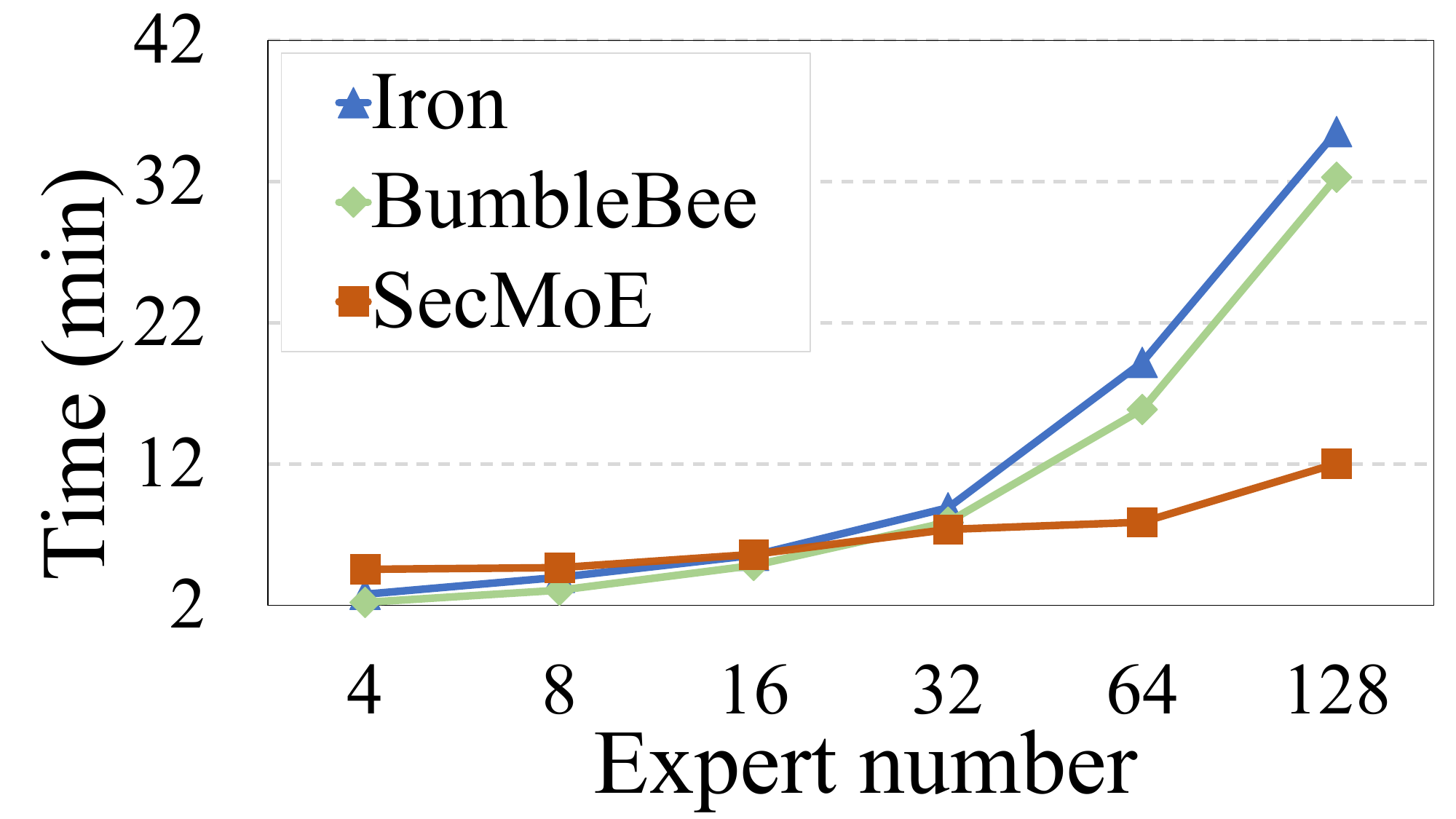}
  \end{subfigure}\hfill
  \begin{subfigure}{0.32\textwidth}
    \includegraphics[width=\linewidth]{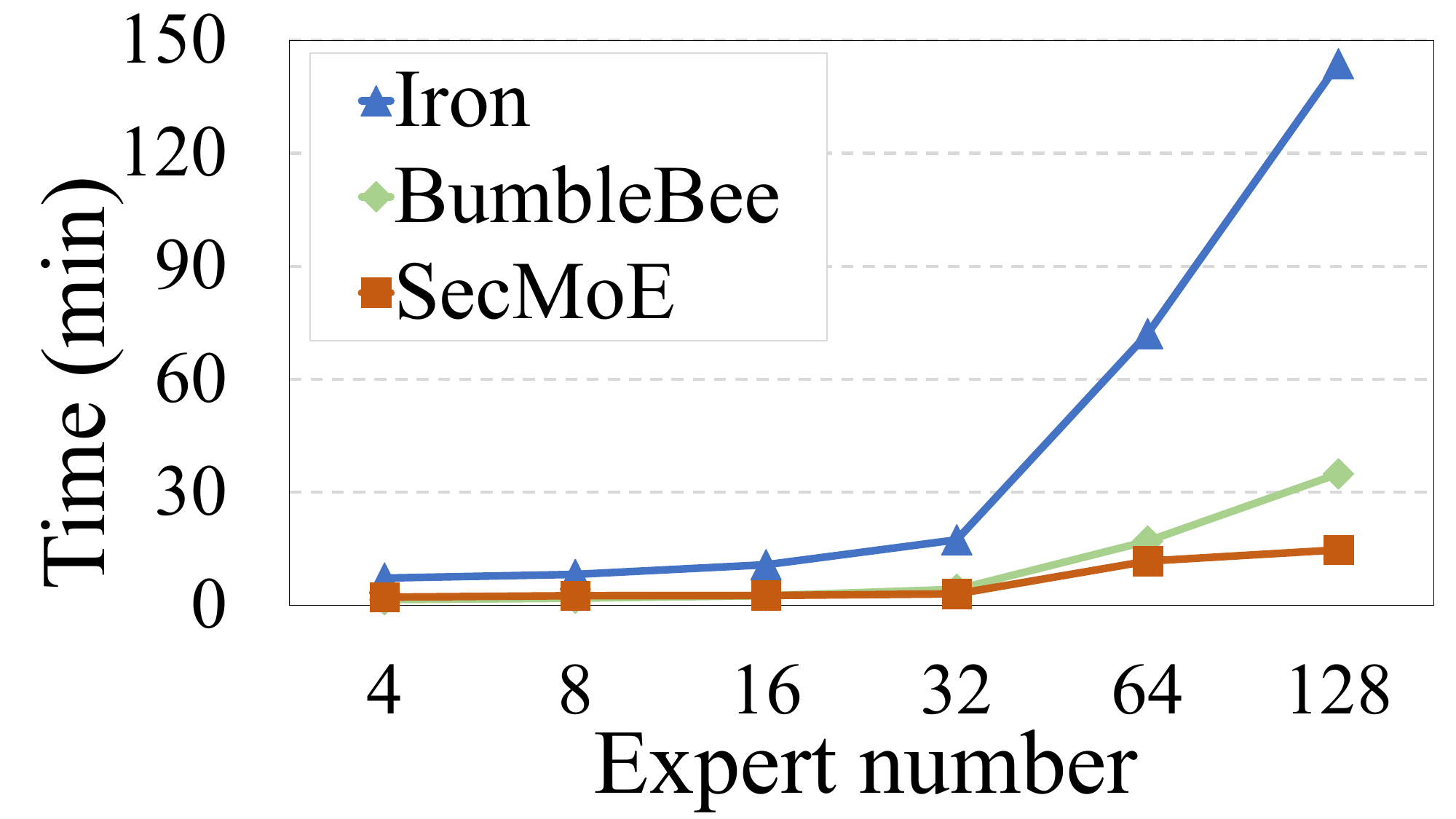}
  \end{subfigure}

  \caption{The private MoE-Small (top) and Switch-Base (bottom) inference for Iron, BumbleBee, and \SecMoE. Begin with communication (MB) and time (s) under LAN (middle) and WAN (right) network settings.}
  \label{fig:23efficiency}
\end{figure*}

\begin{algorithm}[!htbp]
   \caption{Secure GeLU $\Pi_{\text{GeLU}}$}
   \label{alg:secure_gelu}
\begin{algorithmic}[1] 
        \renewcommand{\algorithmicrequire}{\textbf{Input:}}
        \REQUIRE The client $C$ and server $S$ each holds $\ashare{x}_c$ and $\ashare{x}_s$, where $x=\ashare{x}_c+\ashare{x}_s$. 
        
        \renewcommand{\algorithmicrequire}{\textbf{Output:}}
        \REQUIRE  Client and server obtain $\ashare{y}_c$ and $\ashare{y}_s$ respectively, where $\Pi_{\text{GeLU}}(x)=\ashare{y}_c+\ashare{y}_s$. $c_i,i\in[m_{\text{seg}}]$ is the plaintext polynomial coefficients open to both parties, where $m_{\text{seg}}$ represents rows of coefficients. 

    \STATE $C$ and $S$ compute $\ashare{\text{ind}}^b$ and each obtain boolean vector share $\ashare{\text{ind}}^b_0$ and $\ashare{\text{ind}}^b_1$. $\ashare{\text{ind}}^b$ is represented as $(\Pi_{\text{comp}}\{x<b_1\},...,\Pi_{\text{comp}}\{x<b_{m_{\text{seg}}-1}\})^{\!\top}$.

    \STATE $C$ and $S$ compute $\ashare{c_r}$ and each obtain arithmetic vector shares $\ashare{c_r}_c$ and $\ashare{c_r}_s$. $\ashare{c_r}$ is calculated as $(\Pi_{\text{MUX}}(\ashare{\text{ind}^{\!\top}}^b,c_0),\!...,\Pi_{\text{MUX}}(\ashare{\text{ind}^{\!\top}}^b,c_{n_{\text{seg}}}))=(\ashare{c_r}_0,\,\ashare{c_r}_1,\,\dots,\,\ashare{c_r}_{n_{\text{seg}}}).$

    \STATE $C$ and $S$ compute $\ashare{x^2}:=\Pi_{\text{Mul}}(x,x)$ and obtain $\ashare{x^2}_c, \ashare{x^2}_s$.

    \STATE For our approximation, the coefficient rows are $k=2$, and $C$, $S$ compute $\ashare{y}:=\Pi_{\text{Mul}}(\ashare{x^2},\ashare{c_r}_0)+\Pi_{\text{Mul}}(\ashare{x},\ashare{c_r}_1)+\ashare{c_r}_2$. $C$ and $S$ obtain $\ashare{y}_c, \ashare{y}_s$.
    
 \end{algorithmic}
 \end{algorithm}
 
\paragraph{Secure GeLU Complexity Analysis.}
The GeLU is approximated on $m_{\text{seg}}$ intervals by a quadratic per interval with open coefficient matrix. We then estimate per‑element costs with batch, which scales communication but does not increase rounds. Let $cost_{\text{mul}}$, $cost_{\text{mux}}$, and $cost_{\text{comp}}$ be the communication cost for secure arithmetic multiplication, boolean-arithmetic multiplication, and the cost of secure comparison at fixed bit precision.

In the selection phase, the protocol identifies the interval via $m_{\text{seg}}-1$ parallel secure comparisons. In the compute phase, $n_{\text{seg}}+1$ $\Pi_{\text{MUX}}$ are computed to get ciphertext coefficients. The protocol then computes self-multiplications $\ashare{x^k}$, where $k$ is the power of the polynomial, and performs secure multiplications with the coefficients and shares. The local additive share additions are free. Thus, the per‑element complexity is $\text{Comm.}=(m_{\text{seg}}-1)\,cost_{\text{comp}}\;+\;(n_{\text{seg}}+1)\,cost_{\text{mux}}+m_{\text{seg}}(k-1)^2\,cost_{\text{mul}}.$ This makes the power of every entry uniform and equal to the maximum power among all segments. Under a piecewise quadratic fit, the multiplicative depth is thus capped at two, and the number of secure multiplications per element is fixed.

\paragraph{Further Optimization.}
% 叙述区间异或
To further reduce the number of secure comparisons and minimize communication rounds, we unify the computation $1\{x<b_i\}$, the input comparisons against each breakpoint. Then we reverse lower $\ashare{\text{ind}}_i\oplus \text{True}$ and concatenate upper $\ashare{\text{ind}}_{i+1}$ comparison results to efficiently obtain the intermediate comparisons for intervals $[b_i,b_{i+1}], i\in[m_{\text{seg}}-1]$, reducing communication overhead.

% 前后分段0处不算
On the other hand, in the PPML setting, the server publicly opens the coefficients of the polynomial approximation for the non-linear function. Consequently, during our preprocessing phase, we also open the matrix $C$. When both parties compute the $(\Pi_{\text{MUX}}(\ashare{\text{ind}^{\!\top}}^b,c_i))$, the computation for the zero entries of $C$ can be directly set to 0, which reduces the number of $\Pi_{\text{MUX}}$ to save part of communication.

\subsection{Other Nonlinear Protocols.}
For the Softmax function, given the input $x$, the output of the Softmax function is computed as $\frac{\text{exp}(x_i-max(x))}{\sum \text{exp}(x_j-max(x))}$. Like the previous works~\cite{dong2023puma,lu2023bumblebee,li2024nimbus, kei2025shaft}, we focus on piecewise polynomial approximation and adopt BumbleBee~\cite{lu2023bumblebee} exponential function. Using Taylor series iteration, the power of the polynomial reaches $2^6$, which is not suitable for Select-Then-Compute. And therefore, we keep the original approximation and show the details in the Appendix.

The LayerNorm is denoted as $\text{LayerNorm}(x)=\gamma\cdot (x-(1/d\cdot \sum_j x_j))\cdot (\sum_{j}(x[j]-\mu)^2)^{-1/2}+\beta$, where $\gamma$ and $\beta$ are hyperparameters and $j$ is the index of input array $x$. We also follow the LayerNorm employment in BumbleBee.

\section{Experiments}\label{chap:experiment}
\paragraph{Experiment Setting.}
Our experimental works on the ring $\mathbb{Z}_{2^{64}}$ with a fixed-point precision of $s = 18$. Benchmarks are simulated on two nodes, with each equipped with 64 vCPUs and 128 GB RAM. Network conditions are emulated under the LAN (1 Gbps bandwidth, 0.5 ms latency) and the WAN network (400 Mbps bandwidth, 4 ms latency). The input consists of eight tokens in one batch.

\begin{table}[!h]
  \centering
  \small
  \renewcommand{\arraystretch}{1.1}
  \begin{tabular}{l|ccc ccc}
    \toprule
    \textbf{Model} & \textbf{Parameters} & $d_{\text{model}}$ &
    $d_{\text{ff}}$ & \textbf{Num. Heads} \\ 
    \midrule
    BERT-base  & 110M  & 768 &3072  & 12   \\
    GPT-2 & 117M & 768  &3072  &12 \\
    \midrule
    T5-small & 60M    & 512   & 2048 & 8 \\ 
    T5-Base   & 0.2B    & 768   & 2048 & 12 \\
    \midrule
    MoE-Small & 124M(8e)    & 512   & 2048   & 8 \\
    Switch-Base  & 0.62B(8e)    & 768   & 2048   & 12 \\
    Switch-Base  & 7B(128e)    & 768   & 2048   & 12 \\
    \bottomrule
  \end{tabular}
    
    \caption{Model configurations. $d_{\text{model}}$, $d_{\text{ff}}$, Num. Heads are the dimensions of the hidden layer and model, and the number of transformer attention heads.}
    \label{tab:model-config}
\end{table}

\begin{table*}[!ht]
    \centering
    
    \begin{tabular}{|c|c|c|c|c|c|c|c|c|c|c|}
    \hline
    ~ & \multicolumn{3}{c|}{\makecell{\textbf{Runtime (min)}\\\textbf{in MoE-Small}}}
      & \multicolumn{3}{c|}{\makecell{\textbf{Runtime (min)}\\\textbf{in Switch-Base}}}
      & \multicolumn{4}{c|}{\makecell{\textbf{Communication}\\\textbf{Comparison (GB)}}} \\
    \hline
        ~ & 32e & 64e & 128e & 32e & 64e & 128e & 16e & 32e & 64e & 128e \\ \hline
        ~ & \multicolumn{6}{c|}{LAN Network} & \multicolumn{4}{c|}{MoE-Small}   \\ \hline
        \bfseries Iron & \valp{4.35}{2.1} & \valp{7.58}{3.3} & \valp{12.07}{4.7}
                        & \valp{8.91}{1.2} & \valp{19.2}{2.4} & \valp{35.5}{2.9}
                        & \valp{7.13}{8.9} & \valp{9.44}{11.2} & \valp{17.19}{21.2} & \valp{24.17}{29.4} \\ \hline
        \bfseries BumbleBee & \valp{3.51}{1.7} & \valp{6.51}{2.9} & \valp{9.76}{3.8}
                         & \valp{7.88}{1.0} & \valp{15.8}{2.0} & \valp{32.3}{2.6}
                         & \valp{1.42}{1.8} & \valp{2.04}{2.4} & \valp{3.37}{4.2} & \valp{5.81}{7.1}  \\ \hline
        \bfseries SecMoE & \bfseries 2.04 & \bfseries 2.23 & \bfseries 2.52
                          & \bfseries 7.37 & \bfseries 7.88 & \bfseries 12.1
                          & \bfseries 0.81 & \bfseries 0.84 & \bfseries 0.81 & \bfseries 0.82  \\ \hline
        ~ & \multicolumn{6}{c|}{WAN Network} & \multicolumn{4}{c|}{Switch-Base}   \\ \hline
        \bfseries Iron & \valp{17.40}{5.7} & \valp{35.19}{10.1} & \valp{59.14}{16.1}
                        & \valp{17.40}{5.7} & \valp{72.12}{6.1} & \valp{143.78}{9.7}
                        & \valp{19.81}{9.2} & \valp{28.51}{13.6} & \valp{48.80}{22.4} & \valp{86.68}{29.8}  \\ \hline
        \bfseries BumbleBee & \valp{4.24}{1.3} & \valp{7.59}{2.1} & \valp{13.88}{3.8}
                         & \valp{4.24}{1.3} & \valp{17.05}{1.4} & \valp{34.89}{2.3}
                         & \valp{3.96}{1.8} & \valp{5.99}{2.9} & \valp{9.98}{4.6} & \valp{19.13}{6.6}  \\ \hline
        \bfseries SecMoE & \bfseries 3.04 & \bfseries 3.47 & \bfseries 3.68
                          & \bfseries 3.04 & \bfseries 11.72 & \bfseries 14.73
                          & \bfseries 2.15 & \bfseries 2.09 & \bfseries 2.18 & \bfseries 2.91  \\ \hline
    \end{tabular}
\caption{Comparing the runtime (min) and communication (GB) costs of Iron, BumbleBee, and \SecMoE under various expert settings and networks.}
\label{tab:effi_comp}
\end{table*}

\paragraph{Metrics.} 
We assess the efficiency of \SecMoE by measuring the running time and communication cost. We focus on evaluating the end-to-end performance and do not distinguish between offline and online costs. The online cost is significantly lower compared with the offline cost. Therefore, it is unfair to only compare the online performance. We choose Accuracy (ACC) and the Matthews Correlation Coefficient (MCC) metric, which will be detailed when the averaged results are presented in Table~\ref{acc}. 

\paragraph{Model and Datasets.}
Table~\ref{tab:model-config} summarizes the model configurations used in our study. To obtain a controlled MoE setting, we construct MoE‑small by starting from T5‑small and replacing each dense FFN with sparse MoE blocks equipped with $N_{\text{exp}}$ experts, keeping the backbone $d_{\text{model}}$ and depth unchanged. We evaluate our method on the two MoE models, MoE‑small and Switch‑Base, $N_{\text{exp}}\in\{8,16,32,128\}$. Compared with the dense baselines, the MoE variants achieve substantially larger parameter capacity (up to 63$\times$), incurring approximately 15.2$\times$ more costs.

\begin{table}[!h]
\centering
\begin{tabular}{lcccc}
\toprule
\textbf{Dataset} & \textbf{Size} & \textbf{Metric} & \textbf{Plaintext} & \textbf{Our work} \\
\midrule
CoLA & 1043  & MCC & 41.0 & 41.0\\
QNLI & 1000  & ACC & 90.3 & 90.2\\
RTE & 277  & ACC & 69.9 & 70.0\\
\bottomrule
\end{tabular}
\caption{Accuracy comparison of plaintext floating-point baseline and \SecMoE.}
\label{acc}
\end{table}

\paragraph{Accuracy Comparison.}
We evaluate accuracy on several validation datasets, including three GLUE~\cite{glue} tasks, the sentence acceptability (CoLA), and natural language inference (RTE and QNLI). The evaluation result is shown in Table~\ref{acc}. Our experiment is built on BumbleBee. This baseline has minor errors due to truncation after multiplication and the conversion from the floating to fixed-point value, which can be fine-tuned to less than 0.05\%. 

% \begin{figure}[!ht]
%     \centering
%     \includegraphics[width=\linewidth]{image/gelu近似.pdf}
%     \caption{Caption}
%     \label{fig:enter-label}
% \end{figure}

\paragraph{Efficiency Comparison.}
The efficiency comparisons for private inference under LAN and WAN settings are shown in Figure~\ref{fig:23efficiency}. We use Iron~\cite{hao2022iron} and Bumblebee~\cite{lu2023bumblebee} as baselines. Some recent works also propose PPML solutions. MPCFormer~\cite{li2022mpcformer} and SHAFT~\cite{kei2025shaft} are implemented on the Crypten~\cite{knott2021crypten} framework, whose preprocessing relies on a trusted dealer, and therefore, we do not compare with them. Nimbus~\cite{li2024nimbus} proposes a protocol allowing the server to send encrypted weights to the client at the preprocessing stage. However, in MoE models, the expanded parameters place high demands on the client's RAM capacity (for up to 30 GB). Therefore, we do not compare with Nimbus and leave the MoE private inference supporting on-demand RAM loading as future work.

Figure~\ref{fig:23efficiency} compares end‑to‑end latency (lower is better) across the number of experts $N_{\text{exp}}\in\{8,16,32,128\}$ for Iron, BumbleBee, and our \SecMoE, where $K_{\text{exp}}=1$ expert is selected. While the two baselines grow rapidly with $N_{\text{exp}}$, \SecMoE remains flat and obtains $5.67\times$ to $11.24\times$ lower latency than Iron and $1.13\times$ to $2.43\times$ over BumbleBee. The advantage widens at larger expert counts, due to \SecMoE’s selection phase reducing redundant experts' computing. We have not tested larger experts (e.g., 256 or more) because of the out-of-memory problem caused by loading model parameters and storing pseudorandom correlations like Beaver Triples\cite{beaver1996correlated}.

Specifically, in the left part of Table~\ref{tab:effi_comp}, SecMoE maintains communication efficiency across both network settings and architectures, with the advantage widening at larger expert counts. In the LAN setting, SecMoE achieves 4.7$\times$ and 3.9$\times$ speedups over Iron and BumbleBee on the MoE‑small model. Under the WAN setting, \SecMoE further extends the advantage due to the low communication in the MoE layers. On MoE‑small with 128 experts, \SecMoE takes 3.68 mins, achieving a 16.1$\times$ speedup compared with Iron. Even against the stronger framework BumbleBee, \SecMoE is 3.8$\times$ faster. On the larger Switch‑Base model, \SecMoE maintains strong gains, achieving up to 9.7$\times$ faster than Iron and 2.3$\times$ faster than BumbleBee. 

For the secure GeLU, we test \SecMoE against BumbleBee for the communication and runtime under LAN and WAN. In the MoE-Small model with 128 experts, GeLU in \SecMoE reduces 44\% of communication and 11\% of LAN runtime compared to BumbleBee. In the Switch-Base model with 128 experts, \SecMoE's GeLU advantage extends to both runtime and bandwidth, achieving a 7.1$\times$ speedup over BumbleBee and reducing communication by 81\%.

\paragraph{Communication Analysis.}
As for the communication shown in the right part of Table~\ref{tab:effi_comp}, \SecMoE's performance remains stable despite the varying number of experts, reducing communication by up to 29.4$\times$ compared to Iron and 6.6$\times$ compared to BumbleBee. For instance, on the MoE‑small model in LAN, \SecMoE's computation grows only 24\% from 32e (2.04 mins) to 128e (2.52 mins), whereas Iron and BumbleBee increase by 178\% over the same range, showing \SecMoE’s robustness to model scaling.

\section{Conclusion}
% To address the inefficiency of deploying MoE using MPC, we propose \SecMoE, a 2‑PC privacy‑preserving inference framework for MoE models. \SecMoE mainly consists of two optimizations via Select-Then-Compute, secure sparse MoE protocol, and secure polynomial selection in nonlinear functions. The former preserves MoE sparsity while avoiding redundant experts' computation, and the latter reduces the communication in piecewise polynomial evaluation by utilizing oblivious selection on extracted parameters. Extensive experiments demonstrate the superiority of \SecMoE. Under 5 expert settings under LAN and WAN, \SecMoE reduces communication by 1.8$\sim$29.8$\times$ and cuts end‑to‑end inference time by up to 16.1$\times$ while preserving the accuracy. Taken together, \SecMoE constitutes, to our knowledge, the first practical protocol for secure MoE inference under the 2-PC semi-honest assumption. In the future, we plan to extend \SecMoE to support dynamic expert routing and load balancing in existing LLM serving pipelines.

To address the inefficiency in MPC-based MoE, we propose \SecMoE, a 2-PC privacy-preserving inference framework. \SecMoE employs two optimizations: secure sparse MoE and secure polynomial selection via Select-Then-Compute, which respectively avoid redundant expert computation and reduce communication in piecewise evaluation. Extensive experiments show that, \SecMoE lowers communication by 1.8$\sim$29.8$\times$ and speeds up end-to-end inference by up to 16.1$\times$. To our knowledge, \SecMoE is the first practical 2-PC protocol for secure MoE inference. 

%In the future, we plan to extend \SecMoE to support dynamic expert routing and load balancing in existing LLM pipelines.

\newpage
\section{Acknowledgements}
The work was supported by the National Natural Science Foundation of China (Grant No. 72442029), National Key R\&D Program of China (2022YFB3102100), Shenzhen Stable Supporting Program (General Project) (GXWD20231130110352002), Humanities and Social Science Fund of Ministry of Education of China (No. 24JZD026), Guangdong Provincial Key Laboratory of Novel Security Intelligence Technologies (2022B1212010005).

\bibliography{aaai2026}

\fi
\setlength{\leftmargini}{20pt}
\makeatletter\def\@listi{\leftmargin\leftmargini \topsep .5em \parsep .5em \itemsep .5em}
\def\@listii{\leftmargin\leftmarginii \labelwidth\leftmarginii \advance\labelwidth-\labelsep \topsep .4em \parsep .4em \itemsep .4em}
\def\@listiii{\leftmargin\leftmarginiii \labelwidth\leftmarginiii \advance\labelwidth-\labelsep \topsep .4em \parsep .4em \itemsep .4em}\makeatother

\setcounter{secnumdepth}{0}
\renewcommand\thesubsection{\arabic{subsection}}
\renewcommand\labelenumi{\thesubsection.\arabic{enumi}}

\newcounter{checksubsection}
\newcounter{checkitem}[checksubsection]

\newcommand{\checksubsection}[1]{%
  \refstepcounter{checksubsection}%
  \paragraph{\arabic{checksubsection}. #1}%
  \setcounter{checkitem}{0}%
}

\newcommand{\checkitem}{%
  \refstepcounter{checkitem}%
  \item[\arabic{checksubsection}.\arabic{checkitem}.]%
}
\newcommand{\question}[2]{\normalcolor\checkitem #1 #2 \color{blue}}
\newcommand{\ifyespoints}[1]{\makebox[0pt][l]{\hspace{-15pt}\normalcolor #1}}

\input{appendix}

\end{document}

%% file: appendix.tex
\section{Appendix}

\subsection{A. Secure GeLU}
We use the following coefficients for our secure GeLU approximation: $c_0=\{0.0,0.0,0.0\}, \\c_1=\{-0.02986296,-0.01380208,-0.00158297\}, \\c_2=\{-0.36497047, -0.23581369,-0.0384032\}, \\c_3=\{0.00485947,  0.50000716,0.3482604\}, \\c_4=\{-0.36491015,1.23575599,-0.03839009\}, \\c_5=\{0.0,1.0,0.0\}$.

\subsection{B. Secure Softmax}
BumbleBee\cite{lu2023bumblebee} computes the exponentiation using the Taylor series with a simple clipping branch:
\begin{equation}
    \exp(x) \approx
\begin{cases}
0 & x < T_{\text{exp}} \\
\left(1 + \dfrac{x}{2^n} \right)^{2^n} & x \in [T_{\text{exp}}, 0]
\end{cases}
\end{equation}
Using the fixed-point precision $f = 18$, BumbleBee sets $\text{Texp} = -13$ as its approximation breakpoint. It sets $n = 6$ to achieve an average error within $2^{-10}$.

\subsection{C. Secure Polynomial Evaluation}

\subsection{D. Related Works}
\label{sec:relatedwork}
\paragraph{Neural Network based private inference}
To solve the problem of privacy protection in two-party neural network inference, some early work explored private inference using secure multi-party computation~\cite{liu2017oblivious,riazi2018chameleon,juvekar2018gazelle,srinivasan2019delphi}. Some hybrid works deploy the private inference with homomorphic encryption and additive secret sharing with Beaver Triples~\cite{beaver1996correlated}, like CrypTFlow2~\cite{rathee2020cryptflow2} and Cheetah~\cite{huang2022cheetah}, achieving significant improvements in efficiency.

\paragraph{Transformer-based private inference}
Extensive research has focused on two-party secure inference mechanisms tailored for the Transformer architecture. In the domain of linear layers, Iron~\cite{hao2022iron} advances the foundational framework established by Cheetah~\cite{huang2022cheetah} by extending the encoding paradigm from matrix-vector multiplication to encompass matrix-matrix multiplication. BumbleBee~\cite{lu2023bumblebee} further compresses multiple output ciphertexts, thereby optimizing the computational complexity and communication overhead. Recently, BOLT~\cite{pang2023bolt} leverages SIMD encoding to enable compact homomorphic cipher packing of linear layers. Nimbus~\cite{li2024nimbus} proposes a client-side outer product protocol to complete model weight encryption in advance in the pre-process phase, reducing the overhead of the online phase. The state-of-the-art SHAFT~\cite{kei2025shaft} achieves constant-round secure softmax and GeLU protocol relying on a trusted dealer, and integrates with the Hugging Face library. Existing PPML schemes adopt the basic Transformer models like BERT and GPT-2, which are sufficient for early Transformers testing but still have a certain gap between the orders of parameters in the plaintext models practically used. In contrast, our work presents a privacy-preserving and efficient MoE inference framework, addressing this critical gap and advancing existing large-scale model inference frameworks toward scalable and sparsity-aware privacy-preserving computation.

\subsection{E. Threat Model}
We consider a static, semi-honest probabilistic polynomial time (PPT) adversary $\mathcal{A}$ in the simulation paradigm~\cite{simulation_model}. In the 2PC setting, the adversary corrupts either the server $S$ or the client $C$ before the function $\Pi_{\mathcal{F}}$ starts. Security is formalized in the real/ideal world: in the $\emph{real}$ execution, $S$ and $C$ run $\Pi_{\mathcal{F}}$ in the presence of $\mathcal{A}$ and an environment $\mathcal{E}$. In the $\emph{ideal}$ execution, the two parties send their inputs to a trusted party that computes the ideal function $\mathcal{F}$. We require that for every real-world adversary $\mathcal{A}$ there exists an ideal-world simulator $\mathsf{Sim}$ such that no PPT environment $\mathcal{E}$ can distinguish the two executions. 

\begin{definition}
A 2-PC protocol $\Pi$ between a server $S$ holding a model description
$\text{W}$ and a client $C$ holding an input $x$ is called a \emph{private inference protocol} if it satisfies the following properties:
$\text{W}$ is represented as $\text{W} = (\text{W}^1,\text{V},\text{W}^2)$ for $k\in\{0,\ldots,N_\text{exp}-1\}$,
and hence
\[
\text{W} = ((\text{W}^1_0,\text{V}_0,\text{W}^2_0),\ldots,
(\text{W}^1_{N_\text{exp}-1},\text{V}_{N_\text{exp}-1},\text{W}^2_{N_\text{exp}-1})).
\]

\begin{itemize}
  \item \textbf{Correctness.} For every $\text{W}$ and $x$, the client's output at the end of $\Pi$ equals the correct calculation $\text{W}(x)$.
  \item \textbf{Privacy.} 
      \emph{Client corruption.} We require a corrupted, semi-honest client to learn nothing about the server's private input $\text{W}$. Formally, there exists a PPT simulator $S_C$ such that $S_C(\text{meta},\text{out})\ \approx^c \text{View}^{\Pi}_C,$ where $\text{View}^{\Pi}_C$ is $C$'s view in the real execution. $\text{out}=\text{W}(x)$, and $\text{meta}$ are the inference output and the public information (e.g., $\text{HE.pp}$, the public key $\text{pk}$, the number of layers $d$ and experts $K$, the size or type of each layer, the activation functions, and the whole model architecture).

      \emph{Server corruption.} We require a corrupted, semi-honest $S$ learn nothing about $C$'s input $x$. Formally, there exists a PPT simulator $S_S$ such that $S_S(\text{meta})\ \approx^c\ \text{View}^{\Pi}_S,$ where $\mathsf{View}^{\Pi}_S$ is $S$'s view in the real execution.
\end{itemize}
\end{definition}

\paragraph{Correctness and Security Proof.}
The correctness and security proof follow the BumbleBee\cite{lu2023bumblebee}. The proof works in a hybrid model, where each component's ideal function 
$\mathcal{F}$ denotes the ideal functionality of the protocol, respectively. In this hybrid model, these functionalities are assumed to be computed correctly and securely.

\paragraph{Security Proof of MoE Protocol.}
The proposed secure Sparse MoE protocol is built upon secure blocks similar to the previous work\cite{lu2023bumblebee}. It guarantees the same security as the layer in BumbleBee. Specifically, in the selection phase, the client and server compute the subprotocols to obtain the selection vector, which is oblivious to both parties, guaranteed by the security of additive secret sharing. Then, the server computes the local homomorphically multiplication with the selection vector and the plaintext model weight without learning any information about the choice. Then the proof of the following protocols is same as the security of the linear protocol in Nimbus\cite{li2024nimbus}.

\paragraph{Security Proof of Nonlinear Protocol.}
In our work, we follow the previous works' subprotocols\cite{lu2023bumblebee,li2024nimbus} to evaluate our piece-wise polynomials without changing the cryptographic protocols. We focus on reordering the execution sequence of subprotocols for piecewise polynomials, improving the online communication efficiency. Specifically, the parameters of the piecewise approximation, the coefficient set $C$ and breakpoints $b$, are fixed and made public in the offline phase. In the online phase, the client and server provide additive shares $\langle x\rangle_c$ and $\langle x\rangle_s$. By the security of the subprotocols and the use of randomness for each invocation, neither party learns any information about the real value of the segment-selection vector. Moreover, we enforce the uniform power of each entry across segments, so neither party learns any information about the chosen polynomial during the secure polynomial evaluation.